\documentclass[prb,twocolumn,showpacs]{revtex4}
\usepackage{amssymb}
\usepackage{amsmath}
\usepackage{epsfig}

\usepackage{soul}
\usepackage{color}

\begin{document}
\title{${\cal PT}-$Symmetric Nonlinear Metamaterials and Zero-Dimensional Systems} 
\author{G. P. Tsironis, N. Lazarides}
\affiliation{
Department of Physics, University of Crete, P. O. Box 2208, 71003 Heraklion,
Greece; \\  $\&$  \\
Institute of Electronic Structure and Laser,
Foundation for Research and Technology-Hellas, \\
P.O. Box 1527, 71110 Heraklion, Greece\\
}


%

%
\begin{abstract}
A one dimensional, parity-time (${\cal PT}$)-symmetric magnetic metamaterial
comprising split-ring resonators having both gain and loss is investigated.
In the linear regime, the transition from the exact to the broken ${\cal PT}$-phase
is determined through the calculation of the eigenfrequency spectrum for two
different configurations; the one with equidistant split-rings and the other 
with the split-rings forming a binary pattern (${\cal PT}$ dimer chain). 
The latter system features a two-band, gapped spectrum with its shape determined 
by the gain/loss coefficient as well as the inter-element coupling.
In the presense of nonlinearity, the ${\cal PT}$ dimer chain
with balanced gain and loss supports nonlinear localized modes in the form of 
novel discrete breathers below the lower branch of the linear spectrum.
These breathers, that can be excited from a weak applied magnetic field by frequency 
chirping, can be subsequently driven solely by the gain for very long times.
The effect of a small imbalance between gain and loss is also considered.
Fundamendal gain-driven breathers occupy both sites of a dimer, while their energy
is almost equally partitioned between the two split-rings, the one with gain
and the other with loss.
We also introduce a model equation for the investigation of classical ${\cal PT}$ 
symmetry in zero dimensions, realized by a simple harmonic oscillator with matched
time-dependent gain and loss that exhibits a transition from oscillatory to diverging 
motion. This behavior is similar to a transition from the exact to the broken ${\cal PT}$ 
phase in higher-dimensional ${\cal PT}-$ symmetric systems. A stability condition 
relating the parameters of the problem is obtained in the case of piecewise 
constant gain/loss function that allows for the construction of a phase diagram with 
alternating stable and unstable regions.
\end{abstract}

\maketitle
\section{Introduction}
The investigation of artificial materials whose properties can be tailored 
has recently attracted a lot of attention. Considerable research effort has 
been invested in the developement of artificial structures that exhibit 
properties not found in nature. Two recent and well known paradigms are the 
metamaterials, that provide full access to all four quadrants of the real 
permittivity - permeability plane \cite{Zheludev2010}, 
and the parity - time (${\cal PT}$) symmetric systems, whose properties rely
on a delicate balance between gain and loss. 
The latter belong to a class of 'synthetic' materials that do not obey separately
the parity ($\cal P$) and time ($\cal T$) symmetries but instead they do exhibit a 
combined ${\cal PT}$ symmetry. The ideas and notions of ${\cal PT}-$symmetric
systems have their roots in quantum mechanics where ${\cal PT}-$symmetric 
Hamiltonians have been studied for many years \cite{Hook2012}. 
The notion of ${\cal PT}$ symmetry has been recently extended to dynamical lattices,
particularly in optics, where photonic lattices combining gain and loss elements 
offer new possibilities for shaping optical beams and pulses.
Soon after the developement of the theory of ${\cal PT}-$symmetric optical lattices
\cite{ElGanainy2007,Makris2008}, the ${\cal PT}-$symmetry breaking was experimentally
observed \cite{Guo2009,Ruter2010,Szameit2011}.
Naturally, such considerations have been also extended to nonlinear lattices
\cite{Dmitriev2010,Miroshnichenko2011} and oligomers \cite{Li2011},
and ${\cal PT}-$related phenomena like unidirectional optical transport 
\cite{Ramezani2010}, unidirectional invisibility \cite{Lin2011}, 
and Talbot effects \cite{Ramezani2012a} were theoretically demonstrated. 
Moreover, it has been shown that optical solitons 
\cite{Dmitriev2010,Suchkov2011,Achilleos2012,Alexeeva2012},
nonlinear modes \cite{Zezyulin2012}, and breathers \cite{Barashenkov2012,Lazarides2013}
may also be supported by ${\cal PT}-$symmetric systems.
Moreover, the application of these ideas in electronic circuits \cite{Schindler2011},
not only provides a platform for testing ${\cal PT}-$related ideas within the 
framework of easily accessible experimental configurations, but also provides 
a direct link to metamaterials whose elements can be modeled with equivalent 
electrical circuits.

Conventional metamaterials comprising resonant metallic elements operate close
to their resonance frequency where unfortunately the losses are untolerably high
and hamper any possibility for their use in device applications.
The pathways to overcome losses are either to replace the metallic parts with 
superconducting ones ({\em superconducting metamaterials}) \cite{Anlage2011},
or to construct {\em active metamaterials} by
incorporating active constituents that provide gain through external energy sourses.
The latter has been recently recognized as a very promising technique for
compensating losses \cite{Boardman2010,Boardman2010b,Si2011}. A particular 
electronic component
that may provide both gain and nonlinearity in a metamaterial is the tunnel (Esaki) 
diode which features a current-voltage characteristic with a negative resistance
part \cite{Esaki1958}. Left-handed transmission lines with successful implementation 
of Esaki diodes have been recently realized \cite{Jiang2011}, although other 
electronic components may be employed as well for loss compensation \cite{Xu2012}.
Thus, the fabrication of ${\cal PT}-$symmetric metamaterials with balanced gain
and loss is feasible with the present technology in the microwaves, combining 
highly conducting split-ring resonators (SRRs) and negative resistance devices
in a way similar to that in electronic circuits \cite{Schindler2011}.
In this prospect, the SRR equivalent circuit parameters and the bias of the negative
resistance device sould be properly adjusted to provide gain and equal amount of 
loss, as well as real eigenfrequencies in a finite frequency range of the gain/loss
parameter. 
 
In the following we present a  one-dimensional, discrete, equivalent circuit model 
for an array of SRRs with alternatingly gain and loss in the two different 
configurations (Section II). In Section III we present linear eigenfrequency spectra 
for systems with small number of SRRs and we obtain the relation that provides the
eigefrequencies for large systems.
It is shown that ${\cal PT}-$symmetric metamaterials undergo spontaneous symmetry
breaking from the exact ${\cal PT}$ phase (real eigenfrequencies) to the broken
${\cal PT}$ phase (at least a pair of complex eigenfrequencies), with variation of the
gain/loss coefficient. 
In Section IV, where nonlinearity becomes important, the generation of long-lived 
nonlinear excitations in the form of discrete breathers (DBs) \cite{Flach2008} is
demonstrated numerically.
These novel gain-driven DBs result by a purely dynamical proccess, through the 
matching of the input power through the gain mechanism and internal loss. 
In Section V we introduce a model ${\cal PT}-$symmetric system in zero dimensions,
realized by a harmonic oscillator with balanced time-periodic gain and loss,
that exhibits extraordinary properties and multiple critical (phase transition) points.
Section VI contains the conclusions. 
\begin{figure}[!h]
\centerline{\epsfig{figure=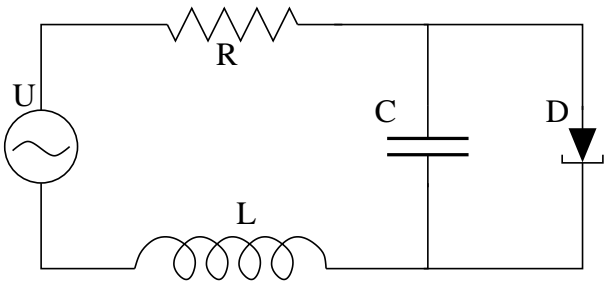,width=80mm}}
\caption{
Electrical equivalent circuit for a split-ring resonator loaded with a tunnel
(Esaki) diode.
}
\label{fig:01}
\end{figure}

\section{Equivalent circuit modelling and Dynamic Equations}
Consider a metallic split-ring resonator (SRR) that can be regarded as an $RLC$ electrical 
circuit featuring Ohmic resistance $R$, inductance $L$, and capacitance $C$.
A tunnel (Esaki) diode is connected in parallel with the capacitance $C$ 
of the SRR (Fig. 1) forming thus a nonlinear metamaterial element with gain.
Esaki diodes exhibit a well defined negative resistance region in their current-voltage
characteristics that has a characteristic $'N'$ shape.
A bias voltage applied to the diode can move its operation point in the 
negative resistance region and then the SRR-diode system gains energy from the source.
\begin{figure}[!t]
\centerline{\epsfig{figure=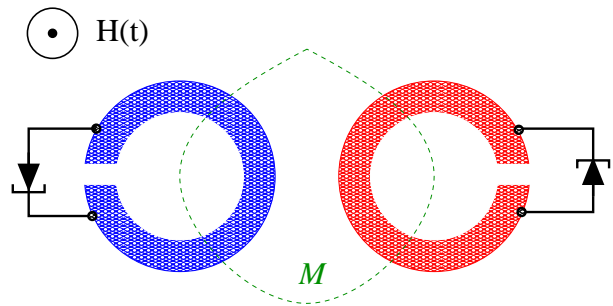,width=80mm}}
\caption{(Color online)
Schematic of a ${\cal PT}-$symmetric metadimer comprising two tunnel
diode-loaded SRRs in an alternating magnetic field $H(t)$. The SRRs are coupled
magnetically through their mutual inductance $M$. Different bias in the diodes
may create a balanced gain/loss structure.
}
\end{figure}

\begin{figure}[!h]
\centerline{\epsfig{figure=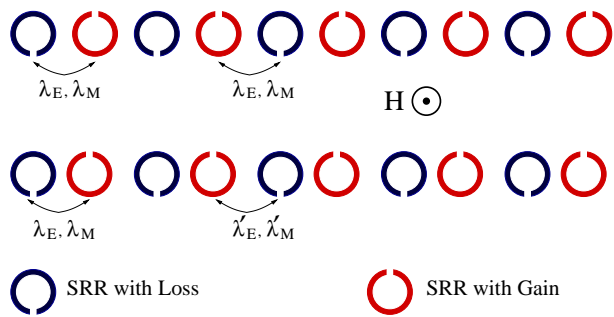,width=80mm}}
\caption{(Color online)
Schematic of a one-dimensional ${\cal PT}-$symmetric metamaterial.
Upper panel: all the split-ring resonators are equidistant.
Lower panel: the separation between split-ring resonators is modulated according 
to a binary pattern (${\cal PT}$ dimer chain).
The applied field is such that its magnetic component is perpendicular to the 
plane of the split-rings.
}
\end{figure}

A metadimer comprising two SRRs loaded with tunnel diodes in an external alternating 
magnetic field is shown in Fig. 2. The equivalent circuit parameters $R$, $C$,
and $L$ of the SRRs and the bias in the diodes have been adjusted so that: 
(i) the two elements have the same eigenfrequencies;
(ii) one of the SRRs has gain while the other has equal amount of loss. 
Then, the pair of SRRs is a ${\cal PT}-$symmetric metadimer that can used 
for the construction of a one-dimensional ${\cal PT}-$symmetric metamaterial, 
which moreover are nonlinear due to the tunnel diodes.
The alternating magnetic field induces an electromotive force (emf) in each SRR 
due to 
Faraday's law which in turn produce currents that couple the SRRs magnetically
through their mutual inductance $M$ (Fig. 2). The coupling strength between SRRs
is rather weak due to the nature of their interaction (magnetoinductive), and has 
been calculated accurately by several authors \cite{Sydoruk2006,Rosanov2011}. 
The SRRs may also be coupled electrically, through the electric dipoles that 
develop in their slits. Thus, in the general case one has to consider both magnetic 
and electric coupling between SRRs. However, for particular relative orientations 
of the SRR slits the magnetic interaction is dominant, while the electric interaction
can be neglected in a first approximation \cite{Hesmer2007,Sersic2009,Feth2010}.   
As can be seen in Fig. 3, the ${\cal PT}-$symmetric metadimers can be arranged in a
one-dimensional lattice in two distinct configurations; one with all the SRRs
equidistant and the other with the SRRs forming a ${\cal PT}$ dimer chain. 

Within the framework of the equivalent circuit model, a set of discrete differential 
equations has been used to describe the dynamics in nonlinear magnetic metamaterials
\cite{Lazarides2006,Molina2009,Lazarides2011,Rosanov2011}. Takining into account the 
binary structure of the ${\cal PT}$ dimer chain, the dynamics of the ${\cal PT}-$symmetric 
metamaterial with balanced gain and loss is governed by the following equations 
which are presented in normalized form  \cite{Lazarides2013}
\begin{figure}[!t]
\centerline{\epsfig{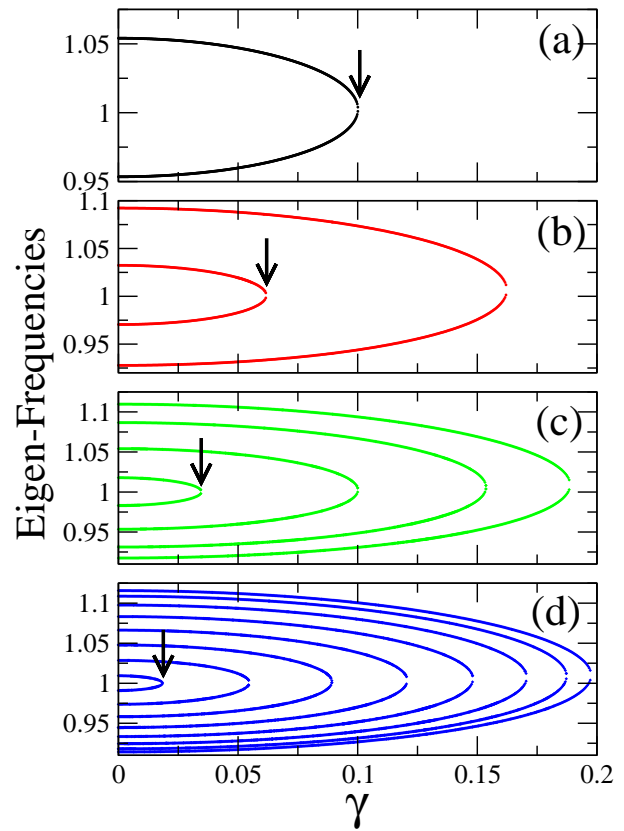}}
\caption{(Color online)
Frequency eigenvalues of the free ${\cal PT}-$symmetric SRR array as a function
of the gain/loss parameter $\gamma$ for $\lambda_E =0$, $\lambda_M =-0.1$,
and (a) $N=2$; (b) $N=4$; (c) $N=8$; (d) $N=16$.
The arrows indicate the critical point $\gamma_c$ for the exact-to-broken
${\cal PT}$ phase transition.
Only the real eigenfrequencies is shown for clarity.
}
\end{figure}
\begin{eqnarray}
\label{1}
   \lambda_M' \ddot{q}_{2n} +\ddot{q}_{2n+1} +\lambda_M \ddot{q}_{2n+2}
 \nonumber \\
  +\lambda_E' q_{2n} +q_{2n+1} +\lambda_E q_{2n+2}
 \nonumber \\
   +\alpha q_{2n+1}^2 +\beta q_{2n+1}^3 +\gamma \dot{q}_{2n+1} 
   =\varepsilon_0 \sin(\Omega \tau) \\ 
\label{2}
   \lambda_M \ddot{q}_{2n-1} +\ddot{q}_{2n} +\lambda_M' \ddot{q}_{2n+1}
 \nonumber \\
   +\lambda_E q_{2n-1} +q_{2n} +\lambda_E' q_{2n+1} 
 \nonumber \\
    +\alpha {q}_{2n}^2 +\beta {q}_{2n}^3 -\gamma \dot{q}_{2n}  
    =\varepsilon_0 \sin(\Omega \tau)
\end{eqnarray}
where
$\lambda_M, \lambda_M'$ and $\lambda_E, \lambda_E'$ are the magnetic and electric
coupling coefficients, respectively, with $\lambda_{E,M} > \lambda_{E,M}'$ and 
$\lambda_{E,M} \lambda_{E,M}' >0$,
$\alpha$ and $\beta$ are dimensionless nonlinear coefficients,
$\gamma$ is the gain/loss coefficient ($\gamma >0$),
$\varepsilon_0$ is the amplitude of the external driving voltage,
while $\Omega$ and $\tau$ are the driving frequency and temporal variable,
respectively, normalized to the inductive-capacitive ($LC$) resonance
frequency $\omega_0$ and inverse $LC$ resonance frequency $\omega_0^{-1}$,
respectively, $\omega_0 =1/\sqrt{L C_0}$ with $C_0$ being the linear capacitance.
\begin{figure}[!t]
\centerline{\epsfig{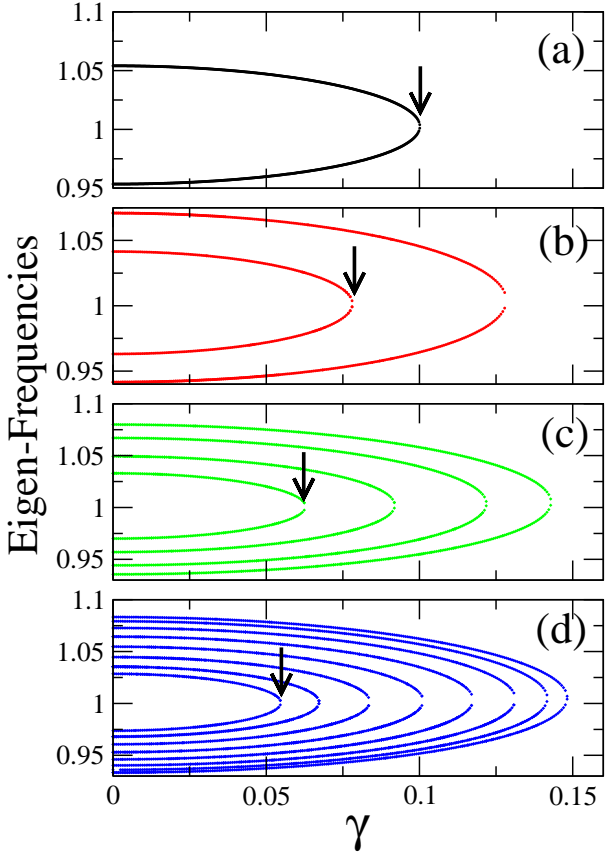}}
\caption{(Color online)
Frequency eigenvalues of the free ${\cal PT}-$symmetric dimer chain as a function
of the gain/loss parameter $\gamma$ for $\lambda_E =\lambda_E' =0$,
$\lambda_M =-0.1$, $\lambda_M =-0.05$,
and (a) $N=2$; (b) $N=4$; (c) $N=8$; (d) $N=16$.
The arrows indicate the critical point $\gamma_c$ for the exact-to-broken
${\cal PT}$ phase transition.
Only the real eigenfrequencies are shown for clarity.
}
\end{figure}
The values selected for the nonlinear coefficients $\alpha=-0.4$, $\beta=0.08$
are typical for a diode and they provide a soft on-site nonlinear potential
for each metamaterial element.
They can be obtained from a Taylor expansion of the capacitance-to -oltage relation
of an equivalent circuit diode model, that gives a very good approximation for
weakly driven systems \cite{Wang2008,Lazarides2011}.

\section{Linear Eigenfrequency Spectra and Critical Point}
In order to obtain the critical value of $\gamma=\gamma_c$ that separates the
exact ${\cal PT}-$phase, where all the eigenvalues are real, from the broken
${\cal PT}-$phase, where at least a pair of eigenvalues is complex, we 
calculate the frequency spectrum. This is a straightforward procedure for systems
with relatively small number of SRRs; the roots of the determinant of the 
linearized Eqs. (\ref{1}) and (\ref{2}) for $\varepsilon_0 =0$ are obtained  
with a root-finding algorithm and then plotted against the gain/loss parameter
$\gamma$. In Figs. 4 and 5, the real eigenfrequencies of ${\cal PT}-$symmetric
metamaterials in both configurations are shown as a function of $\gamma$, while 
the arrows indicate the critical point $\gamma_c$ in each case.
Thus, for $\gamma <\gamma_c$ all eigenvalues are real, while in the opposite case 
at least a pair of eigenvalues has become complex. As we can see from the figures,
for $\gamma > \gamma_c$ more and more eigenfrequency pairs become complex with 
increasing $\gamma$, until they all become complex for a particular value of
$\gamma$.
Moreover, as we can see from an inspection of Figs. 4 and 5, obtained for the 
equidistant SRR configuration and the ${\cal PT}$ dimer chain, respectively,
the value of $\gamma_c$ decreases rapidly with increasing number of SRRs $N$
for the former configuration,
while it tends to a constant value for the latter configuration.  
This can be seen more clearly in Fig. 6, where the critical point $\gamma_c$
is plotted as a function of $N$ for both configurations. For the curves corresponding
to equidistant SRRs (corresponding to two different values of the magnetic coupling
coefficient $\lambda_M$) we see that $\gamma_c$ is smaller for lower magnetic 
coupling coefficient $\lambda_M$.
However, in both curves corresponding to equidistant 
SRRs the value of the critical point $\gamma_c$ tends to zero with increasing $N$.
In contrast, for the ${\cal PT}$ dimer chain configuration, the value of $\gamma_c$
tends to a constant finite value which approximatelly equals the absolute difference 
of the magnetic coupling coefficients $\lambda_M$ and $\lambda_M'$ (see below). 
\begin{figure}[!t]
\centerline{\epsfig{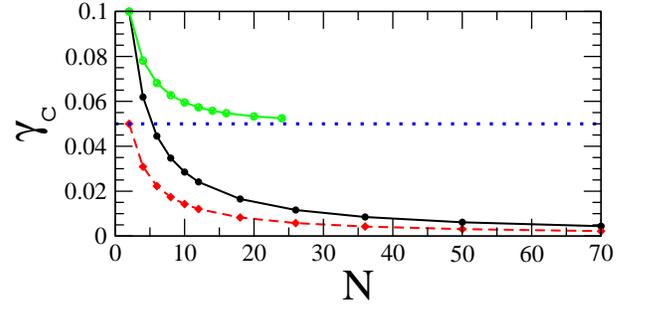}}
\caption{(Color online)
Dependence of the critical gain/loss parameter value $\gamma_c$
on the number of SRRs, $N$, for magnetically coupled SRRs in both the equidistant
and dimer chain configuration. The black squares and the red diamonds have been
calculated for the former configuration with $\lambda_M=-0.1$ and $\lambda_M=-0.05$,
respectively. The green circles have been calculated for the latter configuration with
$\lambda_M=-0.1$, $\lambda_M'=-0.05$. The lines serve as a guide to the eye.
}
\end{figure}

For large systems, we can obtain a condition that determines the critical point $\gamma_c$
as a function of the magnetic coupling constant(s). In the standard way, we substitute 
into the linearized Eqs. (\ref{1}) and (\ref{2}) for $\varepsilon_0 =0$ the trial solutions 
\begin{eqnarray}
 \label{3}
  q_{2n} &=& A \exp[i( 2 n \kappa -\Omega \tau)] , \\
  q_{2n+1}&=& B \exp\{i[ (2 n+1) \kappa -\Omega \tau]\} ,
\end{eqnarray}
where $\kappa$ is the normalized wavevector.
Then, by requesting nontrivial solutions for the resulting stationary problem,
we obtain
\begin{equation}
 \label{4}
  \Omega_\kappa^2 =\left(  -b \pm \sqrt{b^2 -4 a c} \right)/(2 a) ,
\end{equation}
where 
\begin{eqnarray}
\label{5}
  a= 1 -(\lambda_M -\lambda_M')^2 -\mu_\kappa \mu_\kappa' , \\
  b= \gamma^2 -2 \left[ 1 -(\lambda_E -\lambda_E') (\lambda_M -\lambda_M') \right]
   \nonumber \\
                 +\varepsilon_\kappa \mu_\kappa' +\varepsilon_\kappa' \mu_\kappa , \\
  c= 1 -(\lambda_E -\lambda_E')^2 -\varepsilon_\kappa \varepsilon_\kappa' ,
\end{eqnarray}
and $\varepsilon_\kappa = 2 \lambda_E \cos(\kappa)$,
$\varepsilon_\kappa' = 2 \lambda_E' \cos(\kappa)$,
$\mu_\kappa = 2 \lambda_M \cos(\kappa)$, $\mu_\kappa' = 2 \lambda_M' \cos(\kappa)$.
In the following, we neglect the electric coupling between SRRs, i.e.,
$\lambda_E = \lambda_E' =0$, for simplicity. Then, Eq. (\ref{4}) reduces to
\begin{eqnarray}
 \label{9}
  \Omega_\kappa^2 = \frac{2-\gamma^2 \pm \sqrt{\gamma^4 -2 \gamma^2 
               +(\lambda_M -\lambda_M')^2 +\mu_\kappa \mu_\kappa'} }
             {2 (1 -(\lambda_M -\lambda_M')^2 -\mu_\kappa \mu_\kappa')} .
\end{eqnarray}
The condition for having real $\Omega_\kappa$ for any $\kappa$ then reads
\begin{eqnarray}
\label{10}
  \cos^2 (\kappa) \geq \frac{\gamma^2 (2 -\gamma^2) -(\lambda_M -\lambda_M')^2}
                            {4 \lambda_M \lambda_M'} .
\end{eqnarray}
It is easy to see that for
$\lambda_M =\lambda_M'$ corresponding to the equidistant SRR configuration,
the earlier condition cannot be satisfied for all $\kappa$'s
for any positive value of the gain/loss coefficient $\gamma$, implying that a
large ${\cal PT}-$symmetric SRR array (Fig. 3, upper panel) will be in the broken phase.
To the contrary, for $\lambda_M \neq \lambda_M'$, i.e., for a ${\cal PT}$ dimer chain
(Fig. 3, lower panel), the above condition is satisfied for all $\kappa$'s for
$\gamma \leq \gamma_c \simeq |\lambda_M -\lambda_M'|$,  ($\gamma^4 \simeq 0$).
In the exact phase ($\gamma < \gamma_c$), the ${\cal PT}-$symmetric dimer array has
a gapped spectrum with two frequency bands, as shown in Fig. 7.
The width of the gap separating the bands decreases with decreasing
$|\lambda_M -\lambda_M'|$ for fixed $\gamma$. For $\gamma \simeq \gamma_c$
the gap closes,  some frequencies in the spectrum acquire an imaginary part
and the ${\cal PT}$ metamaterial enters into the broken phase.
Note that the gain/loss coefficient $\gamma$ has little effect on the dispersion
curves of the ${\cal PT}$ dimer chain (compare with the dotted curves where 
$\gamma$ is set to zero), as long as the sign in front of $\gamma$ alternates from
one SRR to another. 
\begin{figure}[!t]
\centerline{\epsfig{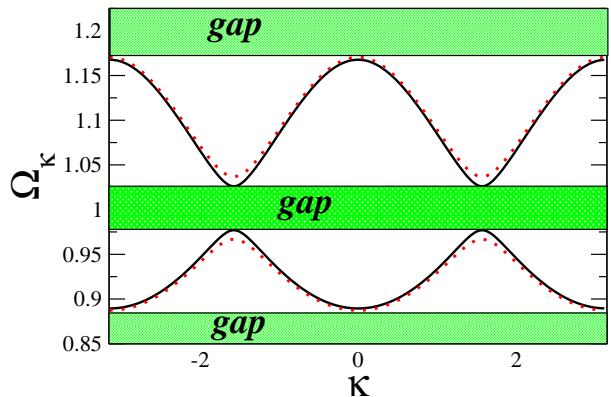}}
\caption{(Color online)
Frequency bands for a ${\cal PT}-$symmetric dimer chain with balanced
gain and loss for $\lambda_M=-0.17$, $\lambda_M' =-0.10$, and
$\gamma =0.05$ (black solid curves); $\gamma =0$ (red dotted curves).
The gaps are indicated in green (dark) color.
}
\end{figure}

\section{Gain-Driven Breather Excitations}
For a gapped linear spectrum, large amplitude linear modes become unstable
in the presence of driving and nonlinearity. If the curvature of the dispersion
curve in the region of such a mode is positive and the lattice potential is
soft, large amplitude modes become unstable with respect to formation of DBs
in the gap below the linear spectrum \cite{Sato2003}.
For the parameters used in Fig. 7, the bottom of the lower band is located
at $\Omega_0 \simeq 0.887$, where the curvature is positive.
The corresponding period at the bottom of the lower band is $T_0 =2\pi/\Omega_0$. 
Moreover, the SRRs are subjected to soft on-site potentials for the selected
values of the nonlinear coefficients $\alpha$ and $\beta$.
Then, DBs can be generated spontaneously by a frequency chirped alternating 
driver; after turning off the driver, the breathers are driven solely by gain.
A similar procedure has been applied succesfuly to lossy nonlinear metamaterials 
with a binary structure \cite{Molina2009,Lazarides2009,Lazarides2010a}
The results are illustrated in Figs. 8 and 9, where the case of a slight imbalance
between gain and loss and its effect on breather generation has been also considered.
In these figures, a density plot of the local energy $E_n$ of a ${\cal PT}-$symmetric
metamaterial is shown on the $n-\tau$ plane for two different values of the driving 
amplitude $\varepsilon_0$.

In the following, the integration of Eqs. (\ref{1}) and (\ref{2}) implemented 
with the boundary condition
\begin{equation}
\label{11}
  q_0 (\tau) = q_{N+1} (\tau) =0 ,
\end{equation}
that accounts for the termination of the structure in finite systems,
is performed with a 4th order Runge-Kutta algorithm with fixed time-step. 
In order to prevent instabilities that will result in divergence of the energy
at particular sites in finite time, we consider a longer dimer chain with total 
number of SRRs $N+2N_\ell$; 
then we replace the gain with equivalent amount of loss at exactly $N_\ell$
SRRs at each end of the extended chain. In other words, we embbed the 
${\cal PT}-$symmetric dimer chain into a geometrically identical lossy
chain, in order to help the excess energy to go smothly away during evolution
living behind stable (or at least very long-lived) breather structures.

We use the following procedure described in detail in Ref. \cite{Lazarides2013}:

\noindent $\bullet$ 
At time $\tau=0$, we start integrating Eqs. (\ref{1}) and (\ref{2}) from 
zero initial state without external driving for $500~T_0 \simeq 3500$ time units (t.u.)
to allow for significant developement of large amplitude modes.

\noindent $\bullet$ 
At time $\tau \simeq 3500$ t.u. the driver is 
switched-on with low-amplitude $\varepsilon_0$ and frequency slightly above $\Omega_0$
($1.01~\Omega_0 \simeq 0.894$). The frequency is then chirped downwards with
time to induce instability for the next $10600$ t.u. ($\sim 1500~T_0$),
until it is well below $\Omega_0$ ($0.997~\Omega_0 \simeq 0.882$).
During that phase, a large number of excitations are generated that move and
strongly interact to each other, eventually merging into a small number of high
amplitude breathers and multi-breathers.

\noindent $\bullet$ 
At time $\tau \simeq 14100$ t.u. (point A on Figs. 8 and 9), the driver is 
switched off and the DBs that have formed are solely driven by the gain.
They continue to interact for some time until they reach an apparently stationary 
state and get trapped at particular sites. The high density segments between
the points A and C in Figs. 8 and 9 precisely depict those gain-driven DB structures.

\noindent $\bullet$  
At time $\tau \sim 440000$ t.u. (point C on Figs. 8 and 9), the gain is replaced 
everywhere by equal amount of loss, and the breathers die out rapidly.
\begin{figure}[!t]
\centerline{\epsfig{figure=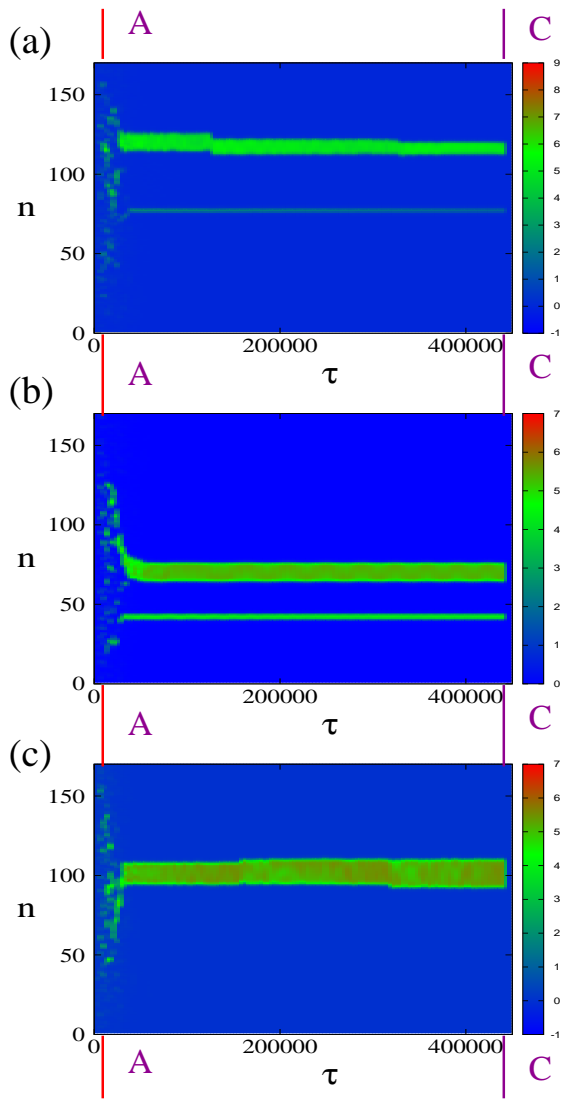,width=80mm}}
\caption{(Color online)
Spatiotemporal evolution of the energy density $E_n$ for a ${\cal PT}-$symmetric
dimer chain with $N=70$, $N_\ell =10$, $\Omega_0 =0.887$, $\gamma=0.002$,
$\lambda_M =-0.17$, $\lambda_M' =-0.10$ ($\lambda_E =\lambda_E' =0$),
$\varepsilon_0 =0.085$, $\gamma=0.002$ and (a) excess loss $0.2\%$;
(b) balanced case; (c) excess gain $0.2\%$.
}
\end{figure}
Note that the above procedure of breather generation is very sensitive to parameter
variations of the external fields. Even though the values of the driving amplitudes 
in Figs. 8 and 9 are rather close (i.e., $\varepsilon_0 =0.085$ and $0.095$, 
respectively), the breather structures as well as their numbers are different.
In Fig. 8(b), in the balanced gain/loss case, we observe two distinct structures that
have been formed that correspond to a relatively high amplitude multi-breather and 
a low amplitude breather. These structures remain stationary during the long time
interval they have been followed ($> 56000~T_0$). In Figs. 8(a) and 8(c), 
gain and loss are not perfectly matched; in Fig. 8(a) loss exceeds gain by a small
amount while in in Fig. 8(c) gain exceeds loss by the same small amount. Notably, 
breather excitations may still be formed through the frequency chirping procedure
in the presence of a small amount of either net gain or net loss.
Indeed, as we may observe comparing Figs. 8(a) and 8(c) with Fig. 8(b), the same 
structures are formed [except the low amplitude breather that is not visible in Fig. 8(c)].
However, both in Figs. 8(a) and 8(c) we can see a slow gradual widening of the high 
amplitude multibreather:
when loss exceeds gain the multibreather losses its energy at a low rate,
with its excited sites that are closer to its end-points gradually falling down to 
a low amplitude state.
Similarly, when gain exceeds loss, the high amplitude multibreather
slowly gains energy and becomes wider. In both cases, breather destruction will take 
place in a time-scale that depends exponentially on the gain/loss imbalance.
Thus, in an experimental situation, where gain/loss balance is only approximate,
it will be still possible to observe breathers at relatively short time-scales. 
Similar observations hold for Fig. 9 as well. In this figure, we observe three 
relatively high amplitude multibreather structures that are formed both in the 
balanced and the imbalanced case.
Here we also observe that instabilities may appear after long time intervals of 
apparently stationary breather evolution as well. 
Whenever this happens, they start moving through the lattice until they get once 
more trapped at different lattice sites. 
In Fig. 9(b) such an instability appears between $~200000-250000$ t.u. for the 
two narrower multibreather structures. The one of them gets trapped a few tenths 
lattice sites away from its previous position, 
while the other (the narrowest) collides and it is absorbed by the wide multibreather
located at $n~50$.
\begin{figure}[!t]
\centerline{\epsfig{figure=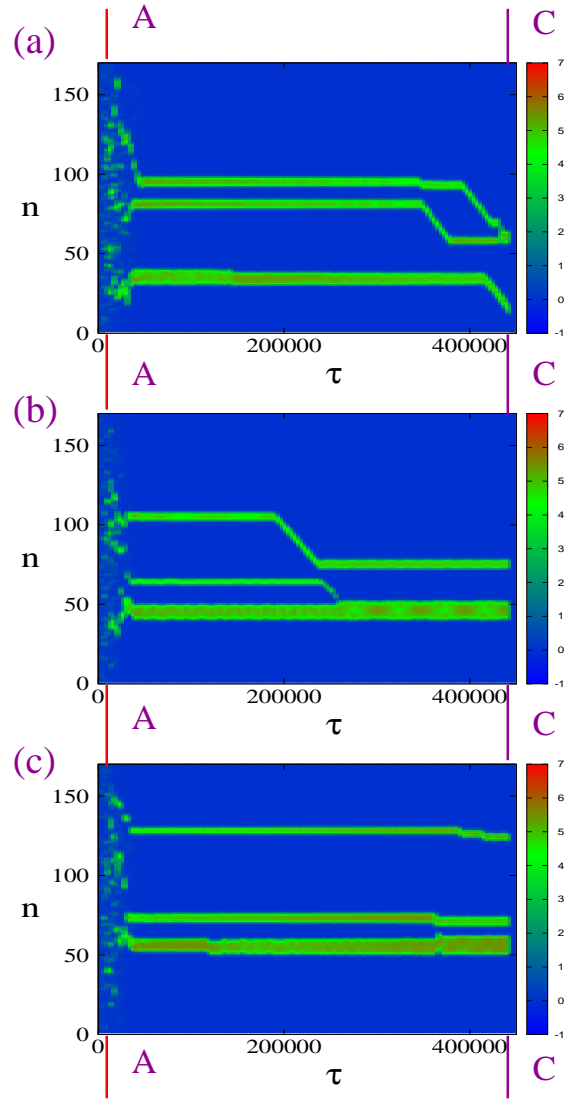,width=80mm}}
\caption{(Color online)
Spatiotemporal evolution of the energy density $E_n$ for a ${\cal PT}-$symmetric
dimer chain with $N=70$, $N_\ell =10$, $\Omega_0 =0.887$, $\gamma=0.002$,
$\lambda_M =-0.17$, $\lambda_M' =-0.10$ ($\lambda_E =\lambda_E' =0$),
$\varepsilon_0 =0.095$, $\gamma=0.002$ and (a) excess loss $0.2\%$;
(b) balanced case; (c) excess gain $0.2\%$.
}
\end{figure}

\section{Classical ${\cal PT}-$Symmetry in Zero Dimensions}
In the classical domain, in all cases of ${\cal PT}-$symmetric systems investigated so
far, the combination of time-reversal and parity symmetries is utilized:
In a given time-evolution if we reverse time while "reflecting" position and momentum,
we must retrace the original path. The parity operation requires that the system is 
extended either in continuous or discrete space in order to be able to perform this 
operation. We show here that this is not necessary and that the features of the
${\cal PT}$ -symmetric systems can be preserved in "zero" dimensions where the parity
symmetry is trivial.

We consider the following  simple harmonic oscillator:
\begin{equation}
  \label{21}
  \ddot{x} +2\theta (t) \dot{x} +\omega_0^2 x=0 ,
\end{equation}
where $x \equiv x(t)$ is the equilibrium dislacement of a mass (the charge in an $RLC$ 
cirquit), $\omega_0$ the resonant oscillation frequency while $\theta (t)$ is a 
time-dependent "damping" term;
both $\omega_0$ and $\theta (t)$ are scaled to the mass (impedance of the circuit).
We take $\theta (t)$ to be periodic with period $T$, viz. $\theta (t+T) =\theta (t)$
while its values may be both positive and negative, i.e., for some part of the period  
the oscillator experiences damping while in the rest of the period time amplification,
or anti-damping. We investigate the osillator evolution after long time and the 
stability of the motion. Instead of addressing a general periodic function $\theta (t)$
we focus on a simple form that makes the problem readily solvable, viz. we take
\begin{equation}
\label{22}           
 \theta (t) = \left\{ \begin{array}{ll}
        +\gamma & \mbox{if $0 \le t<\tau_1$};\\
        -\gamma & \mbox{if $\tau_1 \le t< \tau_2$}.\end{array} \right.            
\end{equation}
where $T=\tau_1 +\tau_2$ and taking the plus (minus) sign in front of the coefficient 
$\gamma$ ($\gamma$ is defined as a positive constant, $\gamma >0$, that may assume 
any value between zero and unity) we have in the first (second) part of the cycle
loss (gain). With this form of piecewise constant
function $\theta (t)$  we can easily solve Eq. (\ref{21}) for the loss (L) segment 
of time duration $\tau_1$ and gain (G) segment of duration $\tau_2$.
The form of Eq. (\ref{22}) permits to view the problem as mapping of the position-velocity 
vector at a given time to the position-velocity vector at a later time;
if in the begining of the gain (loss) period (assuming $t=0$) we have position
and velocity equal to $(x_0 ,~\dot{x}_0)$. Then after the evolution during time $\tau$
($\tau_1$ or $\tau_2$, respectively) we obtain:
\begin{equation}
\label{23}
 \left( \begin{array}{c}
x  \\
\dot{x} \end{array} \right) =  M_{G/L} (\tau)
\left( \begin{array}{c}
x_0  \\
\dot{x_0} \end{array} \right)
\end{equation}
where for gain we have
\begin{eqnarray}
\label{25}
  M_G (\tau_1 ) = 
  \frac{e^{\gamma \tau_1}}{\delta} \times \nonumber \\ 
  \left( \begin{array}{cc}
  \delta \cos\delta \tau_1 -\gamma \sin \delta \tau_1  & \sin \delta \tau_1  \\
  -\omega_0^2 \sin \delta \tau_1  &  \delta \cos\delta \tau_1 +\gamma \sin \delta \tau_1 
  \end{array} \right) 
\end{eqnarray}
and for the loss respectively
\begin{eqnarray}
\label{26}
  M_L (\tau_1 ) = 
  \frac{e^{-\gamma \tau_2}}{\delta} \times \nonumber \\
  \left( \begin{array}{cc}
  \delta \cos\delta \tau_2 +\gamma \sin \delta \tau_2  & \sin \delta \tau_2  \\
  -\omega_0^2 \sin \delta \tau_2  &  \delta \cos\delta \tau_1 -\gamma \sin \delta \tau_2 
  \end{array} \right) 
\end{eqnarray}
where $\delta = \sqrt{\omega_0^2 - \gamma^2}$.  
Using the mapping, we may obtain long time evolution after $N$ periods $T$ as a 
repetitive application of the matrices $M_G (\tau_1)$ and $M_L (\tau_2 )$ to an 
arbitraty initial state $(x_0 ,~\dot{x}_0 )$.  
Since the matrices $M_{L/G} (\tau) e^{\pm \gamma \tau}$ are unimodular, the long time 
evolution with be dominated by the exponential term
$\exp [ N \gamma (\tau_1 -\tau_2 )]$;  this leads trivially to exponential growth
($\tau_1 > \tau_2 $) or exponential decay ($\tau_1 < \tau_2 $).  
As a result we consider the more interesting case with $\tau_1 = \tau_2 =\tau \equiv T/2$;
in this case the gain and loss power is perfectly matched during the period $T$.
The combined propagation matrix after one period (assuming first gain) is simply
$M(T) = M_L (\tau ) M_G (\tau )$, i.e.,`
\begin{equation}
\label{26b}
M (T) = 
\frac{1}{\delta^2}  
\left( \begin{array}{cc}
    M_{11} & M_{12}  \\
    M_{21} & M_{22}  
\end{array} \right) 
\end{equation}
where $M_{ij}$ are given by
\begin{eqnarray}
\label{26.1}
   M_{11} =-\gamma^2 +\omega_0^2 \cos (2\phi) , \\
\label{26.2}
   M_{12} =+2 \sin \phi (\delta \cos \phi + \gamma \sin \phi ) \\
\label{26.3}
   M_{21} =-2 \omega_0^2 \sin \phi (\delta \cos \phi - \gamma \sin \phi ) \\
\label{26.4}
  M_{22} =-\gamma^2 +\omega_0^2 \cos (2\phi) 
\end{eqnarray}
and $\phi = \delta \tau \equiv \delta T/2$.  
The matrix $M$ is clearly also unimodular with eigenvalues $e^{i \mu }$ and 
$e^{- i \mu }$; since the trace of a matrix is invariant we find
\begin{equation}
\label{27}
\cos \mu = \frac{-\gamma^2 +\omega_0^2 \cos 2\phi }{\delta^2} .
\end{equation}

\begin{figure}[!t]
\centerline{\epsfig{figure=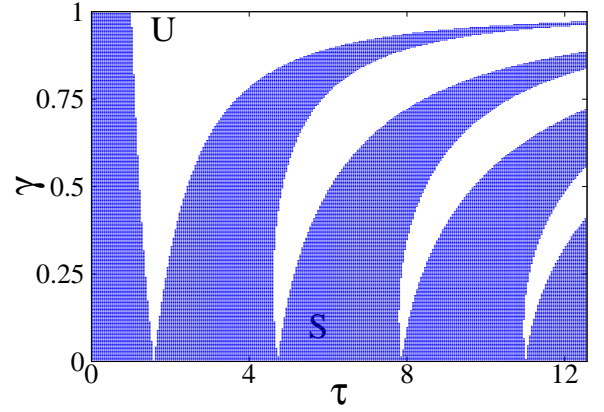,width=80mm}}
\caption{(Color online)
Phase diagram on the $\gamma -\tau$ plane for the harmonic oscillator with
time-dependent "damping" term, obtained from Eq. (\ref{33}).
The stability region is indicated in blue (dark) color.
}
\end{figure}
\noindent {\em Stability Equation.-} Eq. (\ref{27}) can be re written as
\begin{equation}
\label{28}
\cos \mu = \frac{1-B^2\cos 2\phi }{1-B^2}=1+\frac{2B^2 \sin^2 \phi}{1-B^2} 
\end{equation}
with $B=\frac{\omega_0}{\gamma}$.
In order to have stable motion it is necessary that $\vert \cos \mu \vert \le 1$,
or $-1\le 1+\frac{2 B^2 \sin^2 \phi}{1-B^2}  \ge 1$, leading to
\begin{equation}
\label{29}
-2\le \frac{2B^2 \sin^2 \phi}{1-B^2}  \le 1
\end{equation}
Eq. (\ref{29}) has solutions only for $\vert B \vert >1$   
or $\vert \gamma \vert < \omega_0$, for $\omega_0 >0$; in the latter case we find
\begin{equation}
\label{30}
| \cos \phi |  \ge \left| \frac{\gamma}{\omega_0} \right|
\end{equation}
The range thus of allowed values for the angle $\phi$ ($\gamma >0$) is
\begin{eqnarray}
\label{31}
  \frac{\gamma}{\omega_0} \le \cos \phi \le 1 \nonumber \\
  -1 \le \cos \phi   \le  -\frac{\gamma}{\omega_0} 
\end{eqnarray}
We note that there are multiple allowed solutions marked by the the lines
$\cos\phi =\pm \gamma /\omega_0$, i.e., for the three parameters of the problem
$\gamma$, $T = 2 \tau $ and $\omega_0$  the transcedental equation
\begin{eqnarray}
\label{32}
  \cos (\delta \tau)= 
   \left\{ \begin{array}{ll}
     +\frac{\gamma}{\omega_0} & \mbox{for $2n\pi -\pi/2<\delta \tau < 2n\pi +\pi/2$}; \\
     -\frac{\gamma}{\omega_0} & \mbox{for $2\pi n +\pi/2 <\delta \tau < 2\pi n +3\pi/2$} ,
    \end{array} \right.  \nonumber 
\end{eqnarray}
where $n=0, \pm 1, \pm 2, ..$, marks the onset of the transition from stable to
unstable evolution. 
This is the equivalent to the ${\cal PT}$ transition from the exact (stable)
to the broken (unstable) phase in this zero-dimensional problem. 
Using Eq. (\ref{30}) we construct a ${\cal PT}$ phase diagram on the $\tau -\gamma$
plane (Fig. 10), where the blue (dark) color indicates regions of stability.
If we fix one of the parameters, variation of the other drives the oscillator
through alternatingly stable and unstable regions, as can be readily observed
from Fig. 10.
\begin{figure}[!h]
\centerline{\epsfig{figure=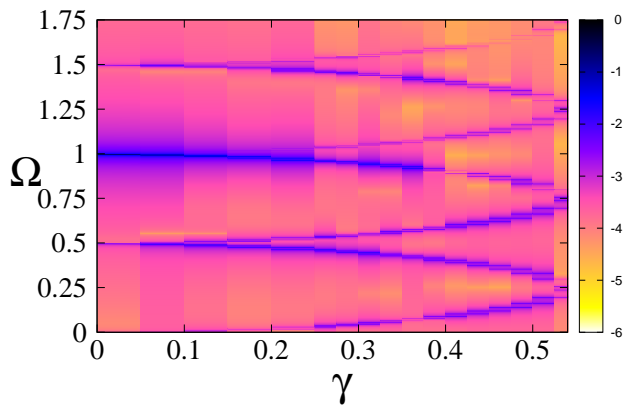,width=90mm}}
\caption{(Color online)
Density plot of the logarithm of the frequency spectra of $x(t)$,
$y=\log_{10}\{PS[x(t)]\}$, as a function of $\gamma$.
The discrete frequency components of the spectra for each value
of $\gamma$ are indicated with blue (darker) color. 
}
\end{figure}

Introducing the frequency $\omega = 2  \pi /T \equiv \pi / \tau$ we define the
reduced parameters $\Omega = \omega / \omega_0 $ and $\tilde{\gamma} =\gamma / \omega_0$. 
Then, the equation that controls the stability regions becomes
\begin{equation}
\label{33}
\left| \cos \left[ \frac{\pi \sqrt{ 1- \gamma^2}}{ \Omega } \right] \right|  = \gamma , 
\end{equation}
where the tilde has been dropped. Eq. (\ref{33}) has solutions for $|\gamma| <1$;
for different reduced frequencies $\Omega$ we obtain different number of solutions
of the earlier equation, the number of the latter icreases with decreasing frequency $\omega$.
Once we have the solutions of Eq. (\ref{33}) we can find the regimes of stability
and instability.
Consider the resonant case $\Omega =1$ where the external frequency of gain/loss
alternation matches the self-frequency of the oscillator.  Besides the trivial solution
at $\gamma=0$, numerical solution of Eq. (\ref{33}) gives $\gamma_1 \approx 0.676$ so 
that the stable region is in the range  $0 \le \gamma \le \gamma_1$.
For $\Omega =0.5$ the corresponding solutions
are $\gamma_1 \approx 0.54$, $\gamma_2 \approx 0.80$, and $\gamma_3 \approx 0.90$ 
with two stability regions, i.e., $0 \le \gamma \le \gamma_1$ and 
$\gamma_2 \le \gamma \le \gamma_3$.

The stable solutions of Eq. (\ref{21}) as a function of time $t$ are in general 
quasi-periodic oscillations whose spectral content varies both with $\tau$ and 
$\gamma$. For an illustration, we consider a particular value of $\tau$, 
i.e., $\tau =2\pi$ ($\Omega=0.5$) for which the boundaries of the stability regions 
have been calculated. For the interval with relatively low values of $\gamma$,
$0 \le \gamma \le \gamma_1 \simeq 0.54$, that is more physically relevant,
we present a density plot of the logarithm of the frequency spectra of $x(t)$,
$y=\log_{10}\{PS[x(t)]\}$, as a function of $\gamma$ (Fig. 11).
In this figure, the discrete frequency components of the spectra for each value
of $\gamma$ are indicated in dark (blue) color.
For $\gamma=0$ and very close to zero, the only frequency appearing in the spectrum
is the eigenfrequency of the oscillator $\Omega \simeq 1$. With increasing $\gamma$,
the frequency components at $\Omega \simeq 1 \pm 0.5$ become more and more important.
At about $\gamma =0.15$ these half-integer frequency components start 
splitting into pairs of frequencies that are symmetric around the half-integer
values. The separation between these pairs increases with further increasing $\gamma$, 
and the frequency components from neighboring pairs come closer and closer together 
until they eventually merge for $\gamma$ approaching its critical value $0.54$. 

\section{Conclusions}
We have investigated theoretically a ${\cal PT}-$symmetric nonlinear metamaterial 
relying on gain and loss. Eigenfrequency spectra for linearized systems of either 
small and large $N$ and two different configurations were calculated and the 
critical points $\gamma_c$ were determined. Large ${\cal PT}-$symmetric metamaterials
with the dimer chain configuration exhibit phase transitions for the exact to the 
broken ${\cal PT}-$symmetry phase, while large ${\cal PT}-$symmetric metamaterials
with the equidistant SRR configuration are always in the broken ${\cal PT}$ phase.   

In the presence of nonlinearity, we have demonstrated numerically the a 
${\cal PT}-$symmetric dimer chain supports localized excitations in the form of
discrete breathers. Breathers are excited by a purely dynamical process, 
with a frequency chirped external magnetic field that induces instability in a 
zero initial state. Subsequently, the nonlinearity focuses energy around points
that have acquired high amplitude leading to the formation of localized structures.
The external field is then switched off and those localized structures are then
solely driven by the gain, while the excess energy lives the system through its lossy
parts at the ends, leading eventually to breather generation. 

Remarkably,
slight imbalance between gain and loss does not destroy the breathers instantly;  
they can still be generated through the frequency chirping procedure and they 
can be regarded as stationary for relatively short time intervals. In the long
term, the imbalanced breathers either gain constantly energy and diverge or
lose constantly energy and vanish. The time-scale for the latter events depends
exponentially on the amount of imbalance.   
    
We have introduced a "zero-dimensional" ${\cal PT}$ system that can be realized
by a harmonic oscillator with a "damping" term that provides balanced gain and loss,
through alternation of the sign of the damping coefficient. 
We consider a piecewise linear gain/loss function and obtain a stability condition,
i.e., a relation between the parameters of the problem. We thus obtain a "phase 
diagram" with parameter regions where oscillatory (stable) motion and diverging
(unstable) motion occur. A crossing of the stability boundary marks the onset of 
a transition from stable to unstable evolution that is equivalent to the ${\cal PT}$
transition from the exact to the broken phase in this zero-dimensional problem.

\section*{acknowledgement}
This research was partially supported by the THALES Project ANEMOS,
co-financed by the European Union (European Social Fund - ESF)
and Greek National Funds through the Operational Program
‘‘Education and Lifelong Learning’’ of the National Strategic Reference Framework 
(NSRF)  ‘‘Investing in knowledge society through the European Social Fund‘‘.

\bibliographystyle{spmpsci}
\bibliography{META13-Library}

\begin{thebibliography}{10}
\providecommand{\url}[1]{{#1}}
\providecommand{\urlprefix}{URL }
\expandafter\ifx\csname urlstyle\endcsname\relax
  \providecommand{\doi}[1]{DOI~\discretionary{}{}{}#1}\else
  \providecommand{\doi}{DOI~\discretionary{}{}{}\begingroup
  \urlstyle{rm}\Url}\fi

\bibitem{Achilleos2012}
Achilleos, V., Kevrekidis, P.G., Frantzeskakis, D.J., Carretero-Gonz{\'a}lez,
  R.: Dark solitons and vortices in pt-symmetric nonlinear media: From
  spontaneous symmetry breaking to nonlinear pt phase transitions.
\newblock Phys. Rev. A \textbf{86}, 013,808 (7pp) (2012)

\bibitem{Alexeeva2012}
Alexeeva, N.V., Barashenkov, I.V., Sukhorukov, A.A., Kivshar, Y.S.: Optical
  solitons in pt-symmetric nonlinear couplers with gain and loss.
\newblock Phys. Rev. A \textbf{85}, 063,837 (13pp) (2012)

\bibitem{Anlage2011}
Anlage, S.M.: The physics and applications of superconducting metamaterials.
\newblock J. Opt. \textbf{13}, 024,001--10 (2011)

\bibitem{Barashenkov2012}
Barashenkov, I.V., Suchkov, S.V., Sukhorukov, A.A., Dmitriev, S.V., Kivshar,
  Y.S.: Breathers in pt-symmetric optical couplers.
\newblock Phys. Rev. A \textbf{86}, 053,809 (12pp) (2012)

\bibitem{Boardman2010}
Boardman, A.D., Grimalsky, V.V., Kivshar, Y.S., Koshevaya, S.V., Lapine, M.,
  Litchinitser, N.M., Malnev, V.N., Noginov, M., Rapoport, Y.G., Shalaev, V.M.:
  Active and tunable metamaterials.
\newblock Laser Photonics Rev. \textbf{5 (2)}, 287--307 (2010)

\bibitem{Boardman2010b}
Boardman, D., King, N., Rapoport, Y.: Circuit model of gain in metamaterials.
\newblock In: C.~Denz, S.~Flach, Y.S. Kivshar (eds.) Nonlinearities in Periodic
  Structures and Metamaterials, \emph{Springer Series in Optical Sciences},
  vol. 150, pp. 259--272. Springer Berlin, Heidelberg (2010)

\bibitem{Dmitriev2010}
Dmitriev, S.V., Sukhorukov, A.A., Kivshar, Y.S.: Binary parity-time-symmetric
  nonlinear lattices with balanced gain and loss.
\newblock Opt. Lett. \textbf{35}, 2976--2978 (2010)

\bibitem{ElGanainy2007}
El-Ganainy, R., Makris, K.G., Christodoulides, D.N., Musslimani, Z.H.: Theory
  of coupled optical pt-symmetric structures.
\newblock Opt. Lett. \textbf{32}, 2632--2634 (2007)

\bibitem{Esaki1958}
Esaki, L.: New phenomenon in narrow germanium $p-n$ junctions.
\newblock Phys. Rep. \textbf{109}, 603--605 (1958)

\bibitem{Feth2010}
Feth, N., K{\"o}nig, M., Husnik, M., Stannigel, K., Niegemann, J., Busch, K.,
  Wegener, M., Linden, S.: Electromagnetic interaction of spit-ring resonators:
  The role of separation and relative orientation.
\newblock Opt. Express \textbf{18}, 6545--6554 (2010)

\bibitem{Flach2008}
Flach, S., Gorbach, A.V.: Discrete breathers - advances in theory and
  applications.
\newblock Phys. Rep. \textbf{467}, 1--116 (2008)

\bibitem{Guo2009}
Guo, A.: Observation of pt-symmetry breaking in complex optical potentials.
\newblock Phys. Rev. Lett. \textbf{103}, 093,902 (2009)

\bibitem{Hesmer2007}
Hesmer, F., Tatartschuk, E., Zhuromskyy, O., Radkovskaya, A.A., Shamonin, M.,
  Hao, T., Stevens, C.J., Faulkner, G., Edwardds, D.J., Shamonina, E.: Coupling
  mechanisms for split-ring resonators:theory and experiment.
\newblock Phys. Stat. Sol. (b) \textbf{244}, 1170--1175 (2007)

\bibitem{Hook2012}
Hook, D.W.: Non-hermittian potentials and real eigenvalues.
\newblock Ann. Phys. (Berlin) \textbf{524 (6-7)}, A106 (2012)

\bibitem{Jiang2011}
Jiang, T., Chang, K., Si, L.M., Ran, L., Xin, H.: Active microwave
  negative-index metamaterial transmission line with gain.
\newblock Phys. Rev. Lett. \textbf{107}, 205,503 (2011)

\bibitem{Lazarides2006}
Lazarides, N., Eleftheriou, M., Tsironis, G.P.: Discrete breathers in nonlinear
  magnetic metamaterials.
\newblock Phys. Rev. Lett. \textbf{97}, 157,406--4 (2006)

\bibitem{Lazarides2009}
Lazarides, N., Molina, M.I., Tsironis, G.P.: Breather induction by modulational
  instability in binary metamaterials.
\newblock Acta Physica Polonica A \textbf{116 (4)}, 635--637 (2009)

\bibitem{Lazarides2010a}
Lazarides, N., Molina, M.I., Tsironis, G.P.: Breathers in one-dimensional
  binary metamaterial models.
\newblock Physica B \textbf{405}, 3007--3011 (2010)

\bibitem{Lazarides2011}
Lazarides, N., Paltoglou, V., Tsironis, G.P.: Nonlinear magnetoinductive
  transmission lines.
\newblock Int. J. Bifurcation Chaos \textbf{21}, 2147--2156 (2011)

\bibitem{Lazarides2013}
Lazarides, N., Tsironis, G.P.: Gain-driven discrete breathers in pt-symmetric
  nonlinear metamaterials.
\newblock Phys. Rev. Lett. \textbf{110}, 053,901 (5pp) (2013)

\bibitem{Li2011}
Li, K., Kevrekidis, P.G.: Pt-symmetric oligomers: Analytical solutions, linear
  stability, and nonlinear dynamics.
\newblock Phys. Rev. E \textbf{83}, 066,608 (7pp) (2011)

\bibitem{Lin2011}
Lin, Z., Ramezani, H., Eichekraut, T., Kottos, T., Cao, H., Christodoulides,
  D.N.: Unidirectional invisibility induced by pt-symmetric periodic
  structures.
\newblock Phys. Rev. Lett. \textbf{106}, 213,901 (4pp) (2011)

\bibitem{Makris2008}
Makris, K.G., El-Ganainy, R., Christodoulides, D.N., Musslimani, Z.H.: Beam
  dynamics in pt-symmetric optical lattices.
\newblock Phys. Rev. Lett. \textbf{100}, 103,904 (2008)

\bibitem{Miroshnichenko2011}
Miroshnichenko, A.E., Malomed, B.A., Kivshar, Y.S.: Nonlinearly pt-symmetric
  systems: Spontaneous symmetry breaking and transmission resonances.
\newblock Phys. Rev. A \textbf{84}, 012,123 (4pp) (2011)

\bibitem{Molina2009}
Molina, M.I., Lazarides, N., Tsironis, G.P.: Bulk and surface magnetoinductive
  breathers in binary metamaterials.
\newblock Phys. Rev. E \textbf{80}, 046,605 (2009)

\bibitem{Ramezani2012a}
Ramezani, H., Christodoulides, D.N., Kovanis, V., Vitebskiy, I., Kottos, T.:
  Pt-symmetric talbot effects.
\newblock Phys. Rev. Lett. \textbf{109}, 033,902 (2012)

\bibitem{Ramezani2010}
Ramezani, H., Kottos, T., El-Ganainy, R., Christodoulides, D.N.: Unidirectional
  nonlinear pt-symmetric optical structures.
\newblock Phys. Rev. A \textbf{82}, 043,803 (6pp) (2010)

\bibitem{Rosanov2011}
Rosanov, N.N., Vysotina, N.V., Shatsev, A.N., Shadrivov, I.V., Powell, D.A.,
  Kivshar, Y.S.: Discrete dissipative localized modes in nonlinear magnetic
  metamaterials.
\newblock Opt. Express \textbf{19}, 26,500 (2011)

\bibitem{Ruter2010}
R{\"u}ter, C.E., Makris, K.G., El-Ganainy, R., Christodoulides, D.N., Segev,
  M., Kip, D.: Observation of parity–time symmetry in optics.
\newblock Nature Physics \textbf{6}, 192-- (2010)

\bibitem{Sato2003}
Sato, M., Hubbard, B.E., Sievers, A.J., Ilic, B., Czaplewski, D.A., Graighead,
  H.G.: Observation of locked intrinsic localized vibrational modes in a
  micromechanical oscillator array.
\newblock Phys. Rev. Lett. \textbf{90}, 044,102 (4pp) (2003)

\bibitem{Schindler2011}
Schindler, J., Li, A., Zheng, M.C., Ellis, F.M., Kottos, T.: Experimental study
  of active lrc circuits with pt symmetries.
\newblock Phys. Rev. A \textbf{84}, 040,101(R) (2011)

\bibitem{Sersic2009}
Sersi\'c, I., Frimmer, M., Verhagen, E., Koenderink, A.F.: Electric and
  magnetic dipole coupling in near-infrared split-ring metamaterial arrays.
\newblock Phys. Rev. Lett. \textbf{103}, 213,902 (2009)

\bibitem{Si2011}
Si, L.M., Jiang, T., Chang, K., T.-C.Chen, Lv, X., Ran, L., Xin, H.: Active
  microwave metamaterials incorporating ideal gain devices.
\newblock Materials \textbf{4}, 73--83 (2011)

\bibitem{Suchkov2011}
Suchkov, S.V., Malomed, B.A., Dmitriev, S.V., Kivshar, Y.S.: Solitons in a
  chain of parity-time invariant dimers.
\newblock Phys. Rev. A \textbf{84}, 046,609 (2011)

\bibitem{Sydoruk2006}
Sydoruk, O., Radkovskaya, A., Zhuromskyy, O., Shamonina, E., Shamonin, M.,
  Stevens, C., Faulkner, G., Edwards, D., Solymar, L.: Tailoring the near-field
  guiding properties of magnetic metamaterials with two resonant elements per
  unit cell.
\newblock Phys. Rev. B \textbf{73}, 224,406 (2006)

\bibitem{Szameit2011}
Szameit, A., Rechtsman, M.C., Bahat-Treidel, O., Segev, M.: Pt-symmetry in
  honeycomb photonic lattices.
\newblock Phys. Rev. A \textbf{84}, 021,806(R) (2011)

\bibitem{Wang2008}
Wang, B., Zhou, J., Koschny, T., Soukoulis, C.M.: Nonlinear properties of
  split-ring resonators.
\newblock Opt. Express \textbf{16}, 16,058 (2008)

\bibitem{Xu2012}
Xu, W., Padilla, W.J., Sonkusale, S.: Loss compensation in metamaterials
  through embedding of active transistor based negative differential resistance
  circuits.
\newblock Opt. Express \textbf{20}, 22,406 (2012)

\bibitem{Zezyulin2012}
Zezyulin, D.A., Konotop, V.V.: Nonlinear modes in finite-dimensional
  pt-symmetric systems.
\newblock Phys. Rev. Lett. \textbf{108}, 213,906 (5pp) (2012)

\bibitem{Zheludev2010}
Zheludev, N.I.: The road ahead for metamaterials.
\newblock Science \textbf{328}, 582--583 (2010)

\end{thebibliography}

\end{document}